\documentclass[fleqn,usenatbib]{mnras}

\usepackage{newtxtext,newtxmath}
\usepackage{booktabs}

\usepackage[T1]{fontenc}
\usepackage{multirow}
\DeclareRobustCommand{\VAN}[3]{#2}
\let\VANthebibliography\thebibliography
\def\thebibliography{\DeclareRobustCommand{\VAN}[3]{##3}\VANthebibliography}

\usepackage{graphicx}	
\usepackage{amsmath}	

\title[A MIGHTEE SFR-Radio Correlation]{A MIGHTEE robust measurement of the star formation rate -- radio correlation}

\author[Charlotte L. Jackson et al.]{Charlotte L. Jackson,$^{1}$\thanks{E-mail: charlotte.jackson@physics.ox.ac.uk}
Matt J. Jarvis,$^{1,2}$
Imogen H. Whittam,$^{1,2}$
James H. Matthews,$^{1}$
\and
Johannes Buchner, $^{3,4}$
Catherine L. Hale,$^{5,1}$
Joel Hamlett,$^{1}$
S. Lyla Jung,$^{1}$ 
Mara Salvato,$^{3,4}$ 
\and
Daniel J. B. Smith,$^{6}$ 
Natalia Stylianou,$^{1}$
and Rohan G. Varadaraj$^{1}$ 
\\
\\
$^{1}$Department of Physics, Astrophysics, University of Oxford, Denys Wilkinson Building, Keble Road, Oxford OX1 3RH, UK\\
$^{2}$Department of Physics and Astronomy, University of the Western Cape, Robert Sobukwe Road, 7535 Bellville, Cape Town, South Africa
\\
$^{3}$ Max Planck Institute for Extraterrestrial Physics, Giessenbachstrasse, 85741 Garching, Germany\\
$^{4}$ Excellence Cluster Universe, Boltzmannstr. 2, D-85748 Garching, Germany\\
$^{5}$Institute for Astronomy, University of Edinburgh, Royal Observatory, Blackford Hill, Edinburgh EH9 3HJ, UK\\
$^{6}$Centre for Astrophysics Research, University of Hertfordshire, Hatfield, Herts AL10 9AB, UK
}
\date{Accepted XXX. Received YYY; in original form ZZZ}

\pubyear{\the\year{}}

\begin{document}
\label{firstpage}
\pagerange{\pageref{firstpage}--\pageref{lastpage}}
\maketitle

\begin{abstract}
Determining the relationship between star-formation rate (SFR) and the radio luminosity ($L_{1.4}$) is critical if we are to trace the star-formation history of the Universe dust-agnostically using current and future radio facilities. However, until now, such work has relied on potentially biased binary classifications of sources to remove contaminating active galactic nuclei (AGN). 
We present a new, statistically-driven methodology for deriving the SFR -- $L_{1.4}$ relation, removing the need for problematic cuts.
We use a Bayesian hierarchical mixture model fit to the radio-detected sources in the deep MIGHTEE COSMOS DR1 catalogue, incorporating the full SFR posterior probability distributions generated by state-of-the-art spectral energy distribution fitting code \texttt{GRAHSP}. This allows us to probabilistically 
determine a mean SFR -- $L_{1.4}$ relation for the SF dominated galaxies, whilst accounting for changing fractions of SF dominated sources across redshift, radio luminosity and stellar mass ranges. 
We find that the SFR -- radio luminosity correlation exhibits a significant dependence on redshift, but a stellar mass dependence that is weaker than
previous studies. Our resultant SFR-radio correlation is $\log_{10}(\text{SFR}/M_{\odot}\,\text{yr}^{-1}) = 0.790\times(\log _{10}(L_{1.4}/\text{W\,Hz}^{-1})-23) + 1.244 \times(1+z)^{0.122} -0.033 \times (\log_{10}(M_*/M_{\odot})-10)$, with an intrinsic scatter of 0.178 dex. We show that this redshift evolution could be explained by a moderate evolution in the radio spectral index of SF galaxies. We attribute the lack of observed strong dependence on stellar mass, compared to recent studies, to the novel statistical approach that does not rely on cuts to remove AGN.
\end{abstract}




\section{Introduction}

Observations of the radio continuum emission from galaxies offer powerful insights into important physical processes, unaffected by intervening dust. In (radio loud) active galactic nuclei (AGN), emission from jets and outflows generally dominate these wavelengths \citep[][]{merloni_measuring_2007,hardcastle_properties_2008,Zakamska2014M,White2017, panessa_origin_2019, Fawcett2020, 2021MNRAS.502.4154R, Jackson2026}, whereas in star-forming galaxies (SFG), it is primarily star formation (SF) activity that gives rise to the radio emission \citep[see][for a review]{1992ARA&A..30..575C}. The connection between SF activity and radio emission arises due to the fact that the most luminous, short-lived stars end in supernovae, and these hugely energetic events accelerate cosmic rays (CRs) to relativistic speeds. The CR electrons and positrons then interact with diffuse galactic magnetic fields, producing synchrotron radiation \citep{draine_physics_2011,2016SAAS...43...85K}. Therefore, 1.4\,GHz radio observations can be used as a delayed tracer of SF activity, and have been shown on many occasions to strongly correlate with the total SFR in a galaxy. 

There has been considerable work over many decades to calibrate the empirical relationship between SFR and 1.4\,GHz radio emission, first often relying on the far-infrared (FIR) emission as a tracer of star formation through the FIR -- radio correlation \citep[e.g.][]{deJong1985,1985ApJ...298L...7H,yun_radio_2001,Appleton2004,Ibar2008,Ivison2010,Jarvis2010,Algera2020, delhaize_vla-cosmos_2017,delvecchio_infrared-radio_2021}, or using dust-corrected emission lines \citep[e.g.][]{duncan_mosdef_2020,seymour_jwst_2026}, and also multi-wavelength estimates of the star-formation rate \citep[e.g.][]{Cram1998,bell_estimating_2003,hodge_radio_2008,2009MNRAS.397.1101G,murphy_calibrating_2011, 2017ApJ...836..185T, 2018MNRAS.475.3010G,Smith2021,2022A&A...664A..83H, cook}. It is generally agreed that the SFR-radio relationship is approximately linear in log-log space, however there is a lack of consensus regarding the universality of any such relation across redshifts, stellar masses and morphology \citep[e.g.][]{delhaize_vla-cosmos_2017,molnar_infrared-radio_2018,molnar_non-linear_2021,delvecchio_infrared-radio_2021,Smith2021,cook}.

There are multiple physical reasons to expect some kind of evolution of the infrared-radio correlation (IRRC) and/or the SFR-radio correlation with redshift and stellar mass ($M_*$). The connection between the infrared (IR) and radio emission in a star forming galaxy hinges on the ability of CR electrons to propagate through the galaxy, and it would be reasonable to assume that this propagation depends on, for example, the mass of that galaxy \citep[see e.g.][]{Smith2021}. Furthermore, the calculation of radio luminosity, $L_{\nu}$, from flux, $S_\nu$, depends on a spectral index parameter such that $S_\nu \propto \nu^{\alpha}$, where $\alpha$ is often taken to be $-0.7$ for non-thermal synchrotron emission \citep{2017A&A...602A...6S, novak_vla-cosmos_2017, kondapally_cosmic_2022}, which is dominant at $1.4\,\text{GHz}$. Any evolution of average radio spectral shape with redshift, either from a steepening from synchrotron cooling \citep[e.g.][]{An2021,An2024,Bait2026}, the increasing contribution from thermal free-free emission \citep[e.g.][]{cr2017} or inverse Compton scattering off of cosmic microwave background photons at higher redshifts \citep{DeZotti2024,whittam2025}, would therefore induce a redshift dependence in the IRRC and the SFR--radio correlations. There are many such potential explanations for an observed redshift evolution in either the IRRC and the SFR--radio correlation, however as the majority of the recent work on this topic has focussed on the IRRC rather than a direct SFR-radio correlation, it is unclear whether any observed evolution is driven by a decoupling of the radio or the IR as a reliable SFR indicator. 

Furthermore, as work in this area has begun to benefit from deep, wide-field radio surveys carried out on instruments such as MeerKAT \citep[][]{Jonas2009, Jonas:2018}, LOFAR \citep[][]{van_haarlem_lofar_2013} and GMRT \citep{swarup,gupta_upgraded_2017}, the issue of cleanly identifying truly SF-dominated sources within catalogues comprised of both faint AGN and SFGs has become increasingly pertinent. Any SFR-radio correlation (or associated IRRC) must of course be calibrated only on sources where it can be safely assumed that the radio emission is a direct result of SF activity. In order to select appropriate samples free from contamination from AGN, where the radio emission is likely dominated by jets and other outflows, selection criteria based on IR colours, optical emission line diagnostics, very-long baseline interferometry or X-ray emission have all been employed in the past to eliminate AGN from radio samples \citep[e.g.][]{Szokoly2004,Donley2012,HerreraRuiz2017,Whittam2022M,Drake2024,Mountrichas2025, Arnaudova2025,peluso_investigating_2020}. 

The use of the IRRC to calibrate a SFR-radio correlation is intrinsically highly assumption-driven. In particular, such an approach assumes that the IR is a perfect tracer of SFR for all galaxies -- which is very unlikely to be true. For example, in low metallicity and high redshift galaxies, the UV emission of a galaxy can be as important to constraining SFR as IR \citep[][]{whitaker2017}, which is unaccounted for in the use of an IRRC. Mass and redshift dependent deviations between the IR and radio emission of a galaxy can also induce evolution in the IRRC that is unrelated to SFR. Furthermore, it is likely that highly obscured AGN, which show no visible-wavelength signs of an AGN component, can heat the dust within both a putative torus and the wider interstellar medium, thus elevating the FIR emission above the level one would expect from SF alone \citep[e.g.][]{Bonfield2011,Symeonidis2016,Maddox2017,Symeonidis2022,Sokol2023}.
However, whilst multi-wavelength spectral energy distribution (SED) fitting \citep[e.g.][]{agnfitter,2022ApJ...927..192Y} can be used to help determine this level of emission for bright, disc-dominated AGN, there remains the tricky subject of identifying those AGN that do not exhibit typical AGN characteristics (e.g. accretion discs, broad line regions) other than a powerful radio jet, often referred to as low-excitation radio galaxies, or LERGs. Past studies have resorted to removing sources that show an excess of radio emission relative to their IR (i.e. have a particularly low $q_{\text{IR}}$) \textit{before then fitting for the IRRC}. 
This methodology runs the risk of being circular, i.e. a given SFR-radio correlation is assumed and sources with "excess radio emission" are removed, and then the correlation is measured, and is also dependant on the (subjective) choice of cut location.  

In this paper, we aim to address some of these critical issues surrounding the calibration of the correlation, and present an improved measurement of the SFR-radio relationship using a rich dataset combining deep radio imaging, exceptional multi-wavelength ancillary data, and state-of-the-art SED fitting. We choose not to employ any form of binary SFG-AGN pre-classification, and instead use a hierarchical Bayesian mixture model to assign sources a likelihood of belonging to a `SF-dominated' or `not SF-dominated' cluster. We also use a SFR derived from detailed SED modelling, using multi-wavelength data from the FUV-FIR. We marginalise over the posterior density distributions of the derived SFRs, include considerations of observational and modelling uncertainties, and are able to identify degeneracies between any mass and redshift evolution of the derived SFR-radio correlation. 

In Section \ref{sec:Obs}, we describe the observational data used in our analysis and outline the SED fitting procedure. In Section \ref{sec:method}, we explain the details of the model used, and in Section \ref{sec:results} present our improved form of the SFR-radio correlation, and discuss any derived evolution with galaxy properties. We assume a \cite{2020A&A...641A...6P} flat $\Lambda$CDM cosmology with $\Omega_m=0.31$, $\Omega_\Lambda=0.69$, $H_0 = 67.66~ 
 \textrm{kms}^{-1}~\textrm{Mpc}^{-1}$. Energies, frequencies and wavelengths are given in the rest-frame.

\section{Observational Data}
\label{sec:Obs} 

\subsection{Multi-wavelength measurements}

The MeerKAT International Tiered GHz Extragalactic Exploration \citep[MIGHTEE; ][]{mattmightee} is a deep radio survey underway using the MeerKAT radio telescope. The survey is being conducted over 20 deg$^{2}$ of sky across four different well-studied fields, namely the Cosmic Evolution Survey (COSMOS), XMM Large-Scale Structure (XMM-LSS), European Large-Area Infrared Space Observatory S1 (ELAIS-S1), and Extended Chandra Deep Field South (CDFS) fields. The resulting images have central rms sensitivities of $\sim1.3$ -- $2.7\mu$Jy beam$^{-1}$ at $\sim1.2$ -- $1.3$ GHz -- see \cite{haledr1} for more details on the Data Release 1 (DR1). We use the host-galaxy associated catalogue \citep{mighteehosts} of the COSMOS field continuum DR1, which contains 20,859 sources in total. Of these, 13,782 lie within the multi-wavelength cross-matching region described in \citet{mighteehosts}, and subsequently 13,155 have a successful host-galaxy association (95.45 per cent) and 13,117 sources have a redshift assigned. Approximately 55 per cent of sources have spectroscopic redshifts from the catalogues of \citet{2015fers.confE..27V, vaccari_spitzer_2026}\footnote{Using the 20th March 2025 version.}, which contain a compilation of measurements from a variety of spectroscopic catalogues e.g. \citet{davies_deep_2025, 2026ApJS..282....6K}. 637 sources have their redshifts taken from the Physics of the Accelerating Universe (PAU) survey \citep{2021MNRAS.501.6103A}. Where neither spectroscopic nor PAU redshifts are available, then photometric redshifts from \citet{natpaper} are used, many of which have their origins in the catalogues of \cite{peterredshifts}.
We do not sample the posteriors of the photometric redshift measurements in our analysis, and refer the interested reader to \citet{ mighteehosts} for a full description of the redshift assignment process of the sample, and \citet{peterredshifts} and \citet{natpaper} for more detail about the photometric redshift methodology.

The Stylianou et al. catalogue contains Visible Infrared Survey Telescope for Astronomy (VISTA) IR imaging from UltraVISTA Data Release 6 \citep[][]{vista}, and optical \textit{grizy} band photometry from DR3 of the HSC-SSP survey \citep[][]{hsc}, as well as \textit{ugriz} data from the Canada-France-Hawaii Telescope Legacy Survey \citep[CFHTLS:][]{cuillandre_introduction_2012}, and is magnitude limited to $K_s = 24.8$. 
We then supplement the catalogue further with mid- and far-IR data from the \textit{Herschel} Extragalactic Legacy Project \citep[HELP;][]{2021MNRAS.507..129S}. Specifically, we use the HELP XID+ photometry data, which have been probabilistically deblended using XID+, as described in Section 4.3 of \citet{2021MNRAS.507..129S}. The advantage of this is that it mitigates issues regarding the combination of low resolution IR data with much higher resolution UV/optical data. We use the \textit{Spitzer} Multiband Image Photometer (MIPS) 24$\mu$m band, the \textit{Herschel} Photoconductor Array Camera and Spectrometer (PACS) 100$\mu$m and 160$\mu$m bands, and the \textit{Herschel} Spectral and Photometric Imaging Receiver (SPIRE) 250$\mu$m, 350$\mu$m, and 500$\mu$m bands. 83 per cent of our sample are detected have MIPS and PACS data, and 87 per cent have detections in all three SPIRE bands. Finally, we access \textit{Spitzer} Infrared Array Camera (IRAC) bands at 3.6 $\mu$m, 4.5 $\mu$m, 5.8 $\mu$m, and 8.0 $\mu$m (available for 70, 71, 72 and 72 per cent of sources, respectively), and near and far ultra-violet (NUV/FUV) Galaxy Evolution Explorer (GALEX) bands at 0.23$\mu$m and 0.15$\mu$m from the Cosmic Evolution Survey (COSMOS) 2020 catalogue \texttt{Farmer} catalogue \citep{cosmos2020}. We use the COSMOS2020 \texttt{Farmer} catalogue, which uses photometry measured by fitting parametric galaxy profiles, taking into account variations in the point-spread function, depth, and source crowding. NUV (FUV) detections are available for 21 (38) per cent of sources due to limitations in GALEX sensitivity, particularly at higher redshifts where the Lyman-break is redshifted through the GALEX filters.
To check the robustness of our work against the effect of non-detections, particularly at the far-IR wavelengths, we also repeated the analysis using only sources with full IR coverage from 1.02-500$\mu$m (7482 sources, 57 per cent of our full sample) and our conclusions do not change. Indeed, the values parametrising our final SFR-radio relationship are consistent within uncertainties. \
\begin{figure}
    \centering
    \includegraphics[width=\linewidth]{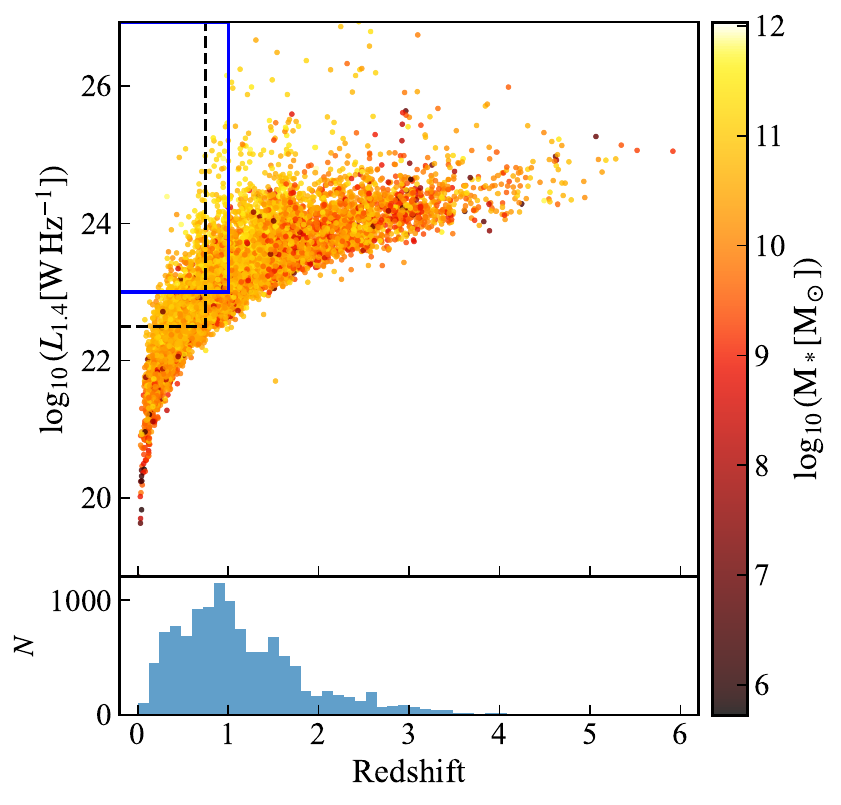}
    \caption{Top: 1.4\,GHz radio luminosity as a function of redshift for the MIGHTEE DR1 (COSMOS field) sample used in this work, coloured by stellar mass. The solid and dashed lines indicate the regions used in the volume-limited test performed in Section \ref{sec:select}. Bottom: Distribution of source redshifts. The median (mean) redshift of the sample is $z=0.99$ ($z=1.16$), and the 16th (84th) percentile is $z=0.46$ ($z=1.76$). 73 sources lie below $z<0.1$, and 72 beyond $z>4$. }
    \label{fig:radz}
\end{figure}
\subsection{SED Fitting}

We conduct multi-band, FUV-FIR SED fitting with the inclusion of a flexible AGN component, using the new SED fitting code \texttt{GRAHSP} \citep[][]{grahsp} to derive galaxy properties such as $M_*$ and SFR. \texttt{GRAHSP} builds upon the foundations of the well-established SED-fitting code \texttt{CIGALE} \citep{cigale, 2022ApJ...927..192Y}, introducing a flexible, empirically validated AGN model component that is comprised of emission from an accretion disc in the form of a power-law continuum, along with broad and narrow lines and a FeII forest, an IR torus component and appropriate galactic and AGN attenuation. This is combined with models of nebular emission and galactic stellar populations, the latter of which makes use of modules developed for \texttt{CIGALE}.

\texttt{GRAHSP} employs a Bayesian approach to constrain model parameters using a nested sampling inference procedure. 
For this study, we adopt a delayed-$\tau$ star formation history (the
\texttt{sfhdelayed} module), a \cite{2003PASP..115..763C} initial mass function with a solar metallicity (Z = 0.02), and \cite{bruzual_stellar_2003} for the simple stellar population (SSP) module. We add nebular emission using the
\texttt{nebular} module \citep{2021A&A...654A.153V}. Dust attenuation is handled by the \texttt{biattenuation} module, which treats the AGN and galaxy components separately. Energy balance is enforced in that absorbed galactic luminosity is remitted in the infrared using the \texttt{galdale2014} module, following \citet{dale_two-parameter_2014}. 
For the AGN components, we use the \texttt{activatepl} module to model the power-law continuum emission
from the accretion disc, the \texttt{activatelines} module to account for both narrow and broad AGN emission lines,
and the \texttt{activategtorus} module to add infrared emission associated with reprocessing of UV light by a dust
torus. A complete description of the SED fitting carried out on the full MIGHTEE DR1 sample will be given in a following paper, Jackson et al. (in prep.). 
\section{Method}
\label{sec:method}

The aim of this work is to agnostically and self-consistently define the SFR–radio relationship using the radio-detected MIGHTEE DR1 sample. To do this, we use star formation rates derived from \texttt{GRAHSP} SED fitting, marginalising over their full posterior probability distributions, and the 1.4\,GHz monochromatic radio luminosities, $L_{1.4}$, calculated from the observed MIGHTEE flux densities as
\begin{equation} \label{eq: lrad}
L_{\textrm{1.4}} = 4\pi d_L^2 S_{\text{obs}}(1+z)^{-\alpha-1} \left( \frac{1.4\text{GHz}}{\nu_{\text{obs}}}\right)^\alpha~ \textrm{W\,Hz$^{-1}$} ,
\end{equation}
\begin{figure*}
\centering 
\includegraphics[width=1\linewidth]{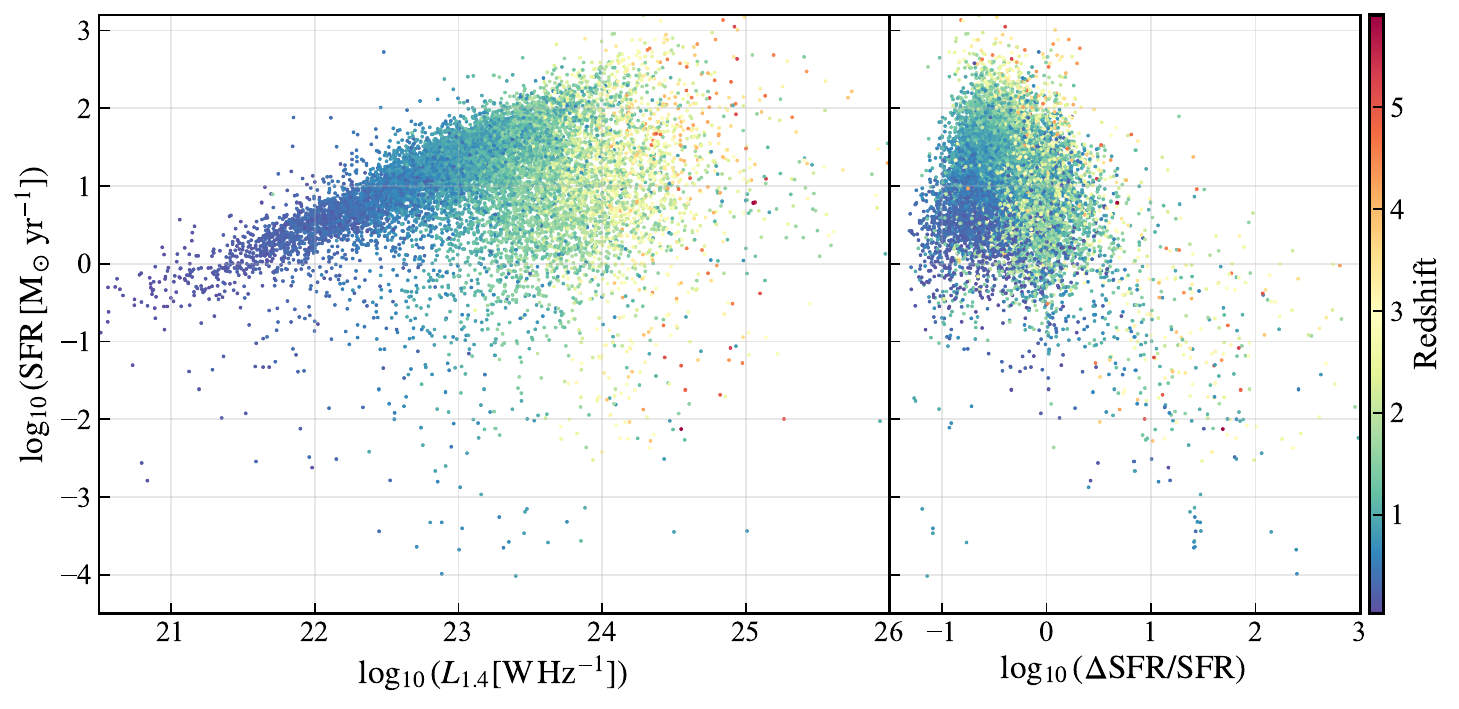} 
\caption{(Left panel) Distribution of intrinsic 1.4\,GHz radio luminosities from MIGHTEE DR1 and SFR from \texttt{GRAHSP} SED fitting for
all sources meeting basic quality cuts described in Section \ref{sec:prep}, coloured by redshift. (Right panel) The SFR shown as a function of the log of the size of the SFR posterior, as defined by the difference between the 84th and 16th percentile values, scaled by the median. Most sources are well-constrained, with 87.6 per cent having $\log_{10}\left(\frac{\Delta\text{SFR}}{\text{SFR}}\right) <0.1$, and 95.7 per cent of sources have $\log_{10}\left(\frac{\Delta\text{SFR}}{\text{SFR}}\right) <0.5$.} 
\label{fig:sfr-loglrad} 
\end{figure*}
where $d_L$ is the luminosity distance, $S_{\textrm{obs}}$ is the observed flux density, $z$ is the redshift, $\nu_{\text{obs}}$ is the observed effective frequency quoted in the MIGHTEE DR1 catalogue and $\alpha$ is the spectral index. Throughout, we assume $S_{\nu} \propto \nu^{\alpha}$ with $\alpha=-0.7$, unless stated otherwise. The analysis in this paper was repeated for $\alpha=-0.75$ and $\alpha=-0.8$, and it was verified that the conclusions were unchanged. For clarity, throughout the paper, $L_{1.4}$ is given in units of $\text{W\,Hz}^{-1}$, SFR is given in units of M$_{\odot}\,\text{yr}^{-1}$, and $M_*$ is given in units of M$_\odot$, unless stated otherwise. Fig.~\ref{fig:radz} shows the redshift versus radio luminosity for the sample used in this paper.

We show in Fig.~\ref{fig:sfr-loglrad} that the distribution of sources in $\log_{10}(L_{1.4})$–$\log_{10}(\text{SFR})$ space produces a clear sequence that corresponds to the expected SFR-radio correlation in SF-dominated sources, alongside a background of radio-excess sources whose radio emission is likely dominated by AGN activity. To preserve our agnosticism and allow our derived relationship to be entirely data-driven, we choose to model both components simultaneously, and infer a SFG-AGN separation probabilistically, rather than removing sources that are likely to be AGN `by hand'. The following sections detail how our model and associated SFR-$L_{1.4}$ relationship is computed. 

\subsection{Dataset Preparation}
\label{sec:prep}

Whilst we wish to remain as agnostic to our sample as possible, it is however necessary to remove sources with unphysical or poorly constrained physical properties. Therefore, we restrict our sample to the redshift range $0.01<z<6$, and remove any sources with an extremely low median SFR, $<10^{-5}~$M$_\odot~{\rm yr}^{-1}$. The exact location of the lower SFR limit does not impact our conclusions. We also quantify the relative size of the SFR posteriors, $\frac{\Delta\text{SFR}}{\text{SFR}}$, as the difference between the 84th and 16th percentile values, scaled by the median, and remove any source with $\log_{10}\left(\frac{\Delta\text{SFR}}{\text{SFR}}\right) > 1$. The vast majority of sources are well-constrained, with 87.6 per cent having $\log_{10}\left(\frac{\Delta\text{SFR}}{\text{SFR}}\right) <0.1$, and 95.7 per cent of sources have $\log_{10}\left(\frac{\Delta\text{SFR}}{\text{SFR}}\right) <0.5$, and the $\log_{10}\left (\frac{\Delta\text{SFR}}{\text{SFR}}\right)$ cut removes 239 sources (1.8 per cent of the sample). The upper and lower redshift bounds remove 9 and 11 sources ($0.07/0.08$ per cent of the total sample) respectively, and 306 sources do not meet SFR median criteria (2.3 per cent). The final sample meeting all quality criteria contains 12083 sources. We discuss the effects of completeness in our sample in Section \ref{sec:select}. 

Then, for numerical stability and to reduce fitted parameter degeneracies, observed quantities are centred around typical values for the population:
\begin{equation}
\label{eq:ref}
L' = \log_{10}(L_{1.4}) - 23, \quad
M' = \log_{10}\left({M_*}\right) - 10.
\end{equation}
Redshift is not centred, as it will ultimately be given a non-linear parameterisation. We want to preserve non-Gaussianity in the SFR posterior distribution whilst improving computational efficiency, so compress each posterior by drawing $R=40$ representative samples (though the overall results are insensitive to this choice), and subsequently assigning each sample a relative weight of $w_{i,r} = 1/R$, for the $i$th source and $r$th sample.

\subsection{Defining the Bayesian Hierarchical Mixture Model}
\label{sec:model}

We model the distribution as a two-component hierarchical Bayesian mixture: a primary SF-dominated relation and a background (non-SF dominated) component. The SF-dominated component is described by 1-dimensional Gaussian scatter in SFR around a mean relation, and the background as a 1-dimensional Gaussian in SFR space. We verified that the results do not change, but equally the overall fit is not improved when the background Gaussian component is allowed to be skewed (i.e. with the skew as an additional free parameter), so we elect to use a standard Gaussian for simplicity.

A crucial element of the model design is how to define the parametric mean of the SF-dominated component, $\mu_{\text{SF}}$, as this ultimately will become our formalised SFR-radio relationship. In its most basic form, this depends only on $\log_{10}(L_{1.4})$ simply given by
\begin{equation}
\label{eq:simple}
\mu_{\text{SF}} = \log_{10}(\mathrm{SFR}\,) = \eta_L L' + \zeta,
\end{equation}
where sampling will optimise for the slope $\eta_L$ and intercept $\zeta$. For clarity, all freely-varying coefficients are labelled $\eta_{\text{parameter}}$ and exponents $\gamma_{\text{parameter}}$. There is a range of literature that suggest that the SFR-radio relationship should include a stellar mass or redshift dependence, or both, \citep[e.g.][]{delhaize_vla-cosmos_2017,2018MNRAS.475.3010G,molnar_non-linear_2021,delvecchio_infrared-radio_2021, Smith2021}. These two properties are inherently observationally linked for flux limited samples, but may also display their own bona fide evolution. Therefore, to test if either of these properties are required to evolve to explain the connection between SFR and radio emission in our sample, we optionally introduce a stellar mass evolution term, $\eta_m M'$, and/or a redshift evolution term, $\eta_z (1+z)^{\gamma_z}$, into the SF-dominated component's mean relation. A constant intercept, $\zeta$, is not necessary when both stellar mass and redshift terms are used additively, as this introduces unreconcilable degeneracies in the fitting. A model with both a stellar mass and redshift dependence would therefore be given by
\begin{equation}
\log_{10}(\text{SFR}) = \eta_L L' + \eta_z (1+z)^{\gamma_z} + \eta_m M'.
\end{equation}

The intrinsic scatter of the SF-dominated component is treated as a free parameter, $\sigma_{\text{int}}$, and is folded into the calculation of the scatter of the SF-dominated Gaussian component, 
\begin{equation}
\sigma_{\text{SF}}^2 = \sigma_{\text{int}}^2 + \eta_L^2 \sigma_{L}^2,
\end{equation}
where $\sigma_L$ is the measurement uncertainty in $\log_{10}(L_{1.4})$, taken directly from the MIGHTEE DR1 catalogue flux errors. 

The background component is modelled as a one-dimensional Gaussian in $\log_{10}(L_{1.4})$ with mean $\mu_{\text{bkg}}$ and scatter $\sigma_{\text{bkg}}$, which are both free parameters.  
The final element of model design is then the prescription of the fraction of sources in the SF-dominated component compared to the background, $f$. The sample used in this work spans wide redshift, stellar mass, and radio luminosity ranges, and therefore it is expected that the population of star forming and active galaxies will not be in constant proportion across the entire parameter space. Failing to account for this would have serious consequences for the robustness of any inferred trends in these properties, as a result of untreated selection effects impacting both populations differently. Therefore, we choose to implement a novel evolving background fraction, and test whether the evolution is best captured by a radio luminosity, redshift, or stellar mass dependance (or some combination of all three). The fraction $f$ is thus parameterised as a logistic function with the form    
\begin{equation}
\centering
 f = \frac{1}{1+\exp\left[-(\phi + \sum_x \psi_{X} X)\right]}  ,
\end{equation}
where $\phi$ is a constant, X is some set of radio luminosity, redshift, and/or stellar mass, and $\psi_{X}$ are the associated coefficients of those variables. 
\begin{table}
\caption{Summary of the models considered in this work. For brevity, the functional form of each model is presented with its dependencies in square brackets, i.e. an additive $\mu_{\text{SF}}$ model that depends on centered stellar mass and redshift is indicated as $\mu_{\text{SF}}[M'+z]$ rather than the full written form, $\log_{10}(\text{SFR}) = \eta_L L' + \eta_z (1+z)^{\gamma_z} + \eta_m M'$. Multiplicative models refer to the form e.g. $\eta \times L'\times M'^{,\gamma_m}\times z^{\gamma_z}$. All $\mu_{\text{SF}}$ models must depend on $L_{1.4}$, so this is omitted for brevity. }
\centering 
\begin{tabular}{ll} 
\hline
\multicolumn{2}{c}{SF-dominated Mean Description, $\mu_{\text{SF}}$} \\
\hline
Functional form& Description\\ 
\hline 
$\mu_{\text{SF}}[\text{base}]$ & $L_{1.4}$-only baseline \\
$\mu_{\text{SF}}[M'] $& Includes $M_{*}$ term\\ 
$\mu_{\text{SF}}[z]$ & Includes $z$ term \\ 
$\mu_{\text{SF}}[M'+z]$  & Full additive model\\ 
$\mu_{\text{SF}}[\text{mult}:z]$& Multiplicative model including $z$\\ 
$\mu_{\text{SF}}[\text{mult}:M']$ & Multiplicative model including $M_{*}$ \\ 
$\mu_{\text{SF}}[\text{mult}:M'\times z]$  & Full multiplicative model \\ 

\hline
\hline
\multicolumn{2}{c}{Population Fraction Description Denominator, $1/f$} \\
\hline
Functional form & Description\\ 
\hline 
$f[\text{base}]$
 & Constant fraction \\ 
$f[L']$
 & $L_{1.4}$ dependence \\ 
$f[z]$& $z$ dependence\\ 
$f[M']$& $M_{*}$ dependence\\ 
$f[L'$+$z$]
 &  $L_{1.4}$ and $z$ dependence\\ 
$f$[$M'$+$z$]& $M_{*}$ and $z$ dependence\\ 
$f$[$L'$+$M'$]& $L_{1.4}$ and $M_{*}$ dependence\\ 
$f$[$L'$+$z$+$M'$]&$L_{1.4}$, $M_{*}$, and $z$ dependence \\ 
\hline 
\end{tabular}

\label{tab:models}
\end{table}

We test an exhaustive suite of models, comprised of all possible combinations of mean SF-dominated and background fraction parameterisations, summarised in Table \ref{tab:models}. We adopt a `flat-top Gaussian' prior distribution for all free parameters, which is defined as a uniform region centered on $0$ with a half-width of $2$, outside of which there are Gaussian tails of width $0.1$, in normalised units. We choose these priors to combine the non-informativeness of standard uniform priors with improved numerical stability when it comes to sampling, due to the removal of discontinuities in probability space.


\subsection{Likelihood Calculations}
Each source $i$ has an associated set of SFR values, $\text{SFR}_{i,r}$, sampled from its SED-derived posterior distribution, where $r$ runs from 0 to $R=40$. The probability of $\text{SFR}_{i,r}$ being drawn from each model component (i.e. the SF-dominated or background) is given by a Gaussian likelihood, 
\begin{equation}
\log p_{\text{comp}}(\text{SFR}_{i,r}|\theta) = -\log\sigma_{\text{comp},i} - \frac{\log(2\pi)}{2} - \frac{(\text{SFR}_{i,r} -\mu_{\text{comp},i})^2}{2\sigma_{\text{comp},i}^2},
\end{equation}
where $\theta$ denotes the full free parameter set, and $\mu_{\text{comp},i}$ and $\sigma_{\text{comp},i}$ are the corresponding mean and standard deviation of each component (e.g. SF-dominated or background), as given in Section \ref{sec:model}.

The per-source-per-sample probability is then marginalised over $\text{SFR}_{i,r}$ to get a per-source probability of being associated with each component, 
\begin{equation}
\log p_{\text{comp}}(\text{SFR}_{i}|\theta)= \log \left[\sum_{r=1}^R w_{i,r} \times p_{\text{comp}}(\text{SFR}_{i,r}|\theta) \right].
\end{equation}  

The total per-source likelihood is thus computed as a sum of both components, each weighted by the fractional occupation of each component as described by the fraction $f$:
\begin{equation}
 p_{\text{mixture}}(\text{SFR}_{i}|\theta) = f_i  p_{\text{SF}}(\text{SFR}_{i}|\theta) + (1-f_i) p_{\text{bkg}}(\text{SFR}_{i}|\theta).
\end{equation}

The overall log likelihood of a model is then given by the sum over all sources: 
\begin{equation}
\log \mathcal{L}(\theta) = \sum_{i=1}^N  \log p_{\text{mixture}}(\text{SFR}_{i}|\theta),
\end{equation}
where $N$ is the number of sources.


\begin{table*}
\caption{
Comparison of candidate models using the ELPD as calculated by the WAIC and PSIS-LOO-CV. Differences are reported relative to the highest-ranked model. Uncertainties on the ELPDs correspond to the estimated standard errors, while uncertainties on the differences correspond to the standard error of the pairwise difference. Model notation is as defined in Table~\ref{tab:models}. The best performing model has a redshift evolution and weak mass evolution, with a mass and luminosity dependent background fraction.
}
\centering
\begin{tabular}{clcccccc}
\hline
Rank & Model & $k$ & ELPD$_{\text{WAIC}}$ & $\Delta$ELPD$_{\text{WAIC}}$ & ELPD$_{\rm LOO}$ &
$\Delta$ELPD$_{\rm LOO}$ \\
\hline
1 & $\mu_{\rm SF}[M'+z]$; $f[L'+M']$
&
10 & $-9399.5 \pm 141.9$ & $0.0$
&
$-9399.5 \pm 141.9$
&
$0.0$
\\

2 &
$\mu_{\rm SF}[M'+z]$; $f[L'+z+M']$
&
11
&
$-9400.4 \pm 141.9$
&
$0.90 \pm 0.77$
&
$-9400.4 \pm 141.9$
&
$0.89 \pm 0.77$
\\

3 &
$\mu_{\rm SF}[z]$; $f[L'+M']$
&
10
&
$-9404.5 \pm 141.8$
&
$5.02 \pm 4.83$
&
$-9404.5 \pm 141.8$
&
$11.78 \pm 4.83$
\\
4 &
$\mu_{\rm SF}[M']$; $f[L'+M']$
&
11
&
$-9417.9 \pm 141.1$
&
$18.39\pm 7.62$
&
$-9417.86 \pm 141.1$
&
$18.38 \pm 7.63$
\\
5 &
$\mu_{\rm SF}[z]$; $f[L'+z+M']$
&
11
&
$-11404.6 \pm 142.0$
&
$2005.10 \pm 39.64$
&
$-11254.2 \pm 134.1$
&
$1854.70 \pm 35.16$
\\

\hline
\end{tabular}
\label{tab:model_comp}
\end{table*}

Depending on the exact specification of $\mu_{\text{SF}}$ and $f$ used in a given model, $\theta$ can contain up to 13 free parameters. To efficiently explore our large suite of possible models, we first perform an initial model screening using a maximum a posteriori (MAP) estimation from the \texttt{PyMC} \texttt{find\_MAP} function, which acts to maximise the value of \(\log \mathcal{L}(\theta) + \log p(\theta)\), where $p(\theta)$ represents the prior. The function uses the L-BFGS-B algorithm \citep[][]{zhu_algorithm_1997}, and multiple random initialisations per model in order to mitigate convergence to local maxima. The resulting maximum log-likelihood, $\log \hat{\mathcal{L}}(\theta)$, is used to estimate the Akaike Information Criterion \citep[AIC;][]{akaike_factor_1987}, using the point estimates. All possible model configurations are ranked by their AIC score, and the five best (i.e. lowest) scoring model are shortlisted for full posterior sampling.

For those shortlisted models, we perform Bayesian inference using a version of a Hamiltonian Monte-Carlo algorithm called the `No U-Turns Sampler' \citep[\texttt{NUTS};][]{nuts} from the Python package \texttt{PyMC} \citep{abril-pla_pymc_2023}. Each run draws 2000 samples, uses 2000 tuning iterations, samples with 4 chains and has a target acceptance of 0.975. It also allows the posterior probability of a source belonging to each component, $p(\text{comp}|\text{SFR}_{i})$, to be calculated. For the primary SF-dominated component, $p(\text{SF-dominated}|\text{SFR}_{i})$ (referred to herein as $p(\text{SF})$ for clarity) corresponds to the probability of a source's radio luminosity being entirely attributable to star formation given its SFR as informed by our mean relation, and is defined as
\begin{equation}
\label{eq:PSF}
p(\text{SF}) = \frac{f_i p_{\text{SF}}(\text{SFR}_{i})}{f_i p_{\text{SF}}(\text{SFR}_{i}) + (1-f_i)p_{\text{bkg}}(\text{SFR}_{i})}.
\end{equation}

\section{Results and Discussion}
\label{sec:results}

In this section, we present the results of both the model selection procedure, and the sampling performed with the optimal model configuration. We further discuss population evolution as described by our evolving background model, investigate the potential impact of an evolution of spectral index with redshift, and test for possible selection effects using a volume-limited subsample of sources.
\subsection{Model Selection}
\label{sec:model select}
The models that produce the five best AIC scores from the initial MAP screening procedure and thus undergo full posterior sampling are detailed in Table \ref{tab:model_comp}. These highest ranking models have a great deal in common: four out of the top five contain a redshift evolution term in the $\mu_{\text{SF}}$ equation, and all five have both a radio luminosity and stellar mass dependence in the fraction of SFG-AGN. Differences in the top scoring models arise from whether a stellar mass term is needed for the SFR-radio relation, and whether a redshift term is preferred in the population fraction, but it is decisive that the data are best described by models with a redshift-dependent SFR-radio relationship and a SFG-AGN fraction that depends on stellar mass and radio luminosity. In order to robustly compare the possible model configurations, we report the expected log-predictive density (ELPD) calculated from the Widely-Applicable Information Criterion \citep[WAIC;][]{2010arXiv1004.2316W} and the Pareto-Smoothed Importance Sampling Leave-One-Out Cross Validation \citep[PSIS-LOO-CV;][]{vehtari_practical_2017,JMLR:v25:19-556}. The WAIC is similar to the AIC, but generalised to include the full posterior distribution from a Bayesian model, penalising models with greater complexity. PSIS-LOO-CV is an efficient implementation of the more general LOO-CV, which quantifies how well a model performs on unseen data. We implement both of these with \texttt{ArviZ} \citep[][]{2019JOSS....4.1143K}. 

Both the WAIC and PSIS-LOO-CV produce the same rankings, given in Table \ref{tab:model_comp}, preferring overall the model with both a redshift and stellar mass dependence, with the functional form   \begin{equation}
\label{eq:mu}
\log_{10}(\text{SFR}) = \eta_L L' + \eta_z (1+z)^{\gamma_z} + \eta_m M',  
\end{equation}
and an evolving population fraction that goes as:
\begin{equation}
\label{eq:f}
f = \frac{1}{1+\exp\left[-(\phi +\psi_L L'+\psi_m M') \right]}.
\end{equation}
However, there is only a $\sim1\sigma$ preference between the highest and second highest scoring models, meaning they cannot be strongly statistically distinguished. These two models have identical $\mu_{\rm SF}$ forms given by equation~(\ref{eq:mu}), with the only difference between them being simply the addition of an explicit redshift term to the population fraction, $f$. Given that the addition of an independent redshift evolution in the background sources does not significantly improve predictive performance, it suggests that any apparent redshift evolution in the fraction of sources that have their radio emission dominated by SF-related processes is instead induced by the evolution of $L_{1.4}$ and $M_*$, insofar as can be constrained by the data. 
Beyond these two best scoring models, the WAIC or PSIS-LOO-CV ELPD becomes dramatically worse. Both models without an $M_*$ term in $\mu_\text{SF}$ also produce highly degenerate, bimodal posterior distributions. Therefore, due to the preference of the WAIC and PSIS-LOO-CV for the $\mu_{\text{SF}}[M' + z]$,$f[L'+M']$ formulation, we choose to retain this as our primary model. 

Explicitly, the data are best described by a SFR-radio  relation that evolves as a function of both stellar mass and redshift, and the fraction of SF-dominated sources is dictated by a combination of stellar mass and radio luminosity.

\begin{figure*}
    \centering
    \includegraphics[width=1\linewidth]{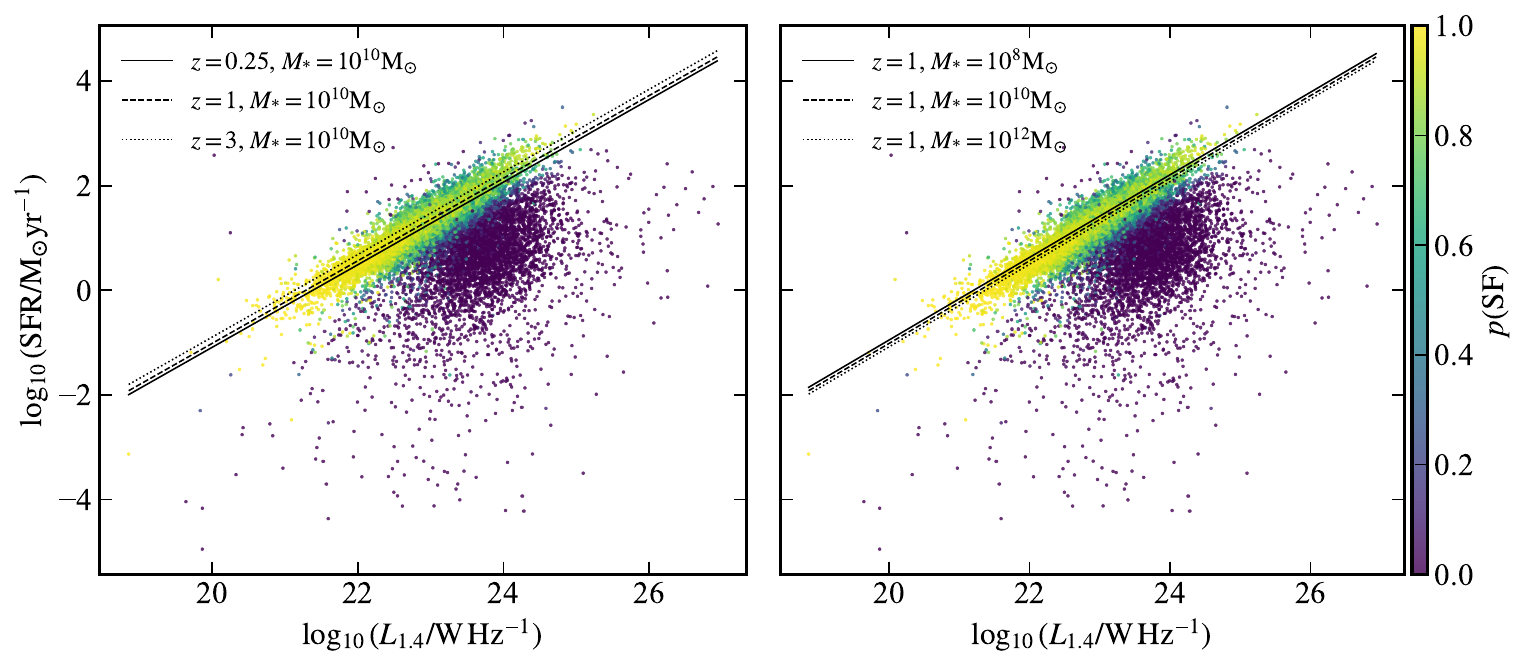}
    \caption{The distribution of sources in $\log_{10}(L_{1.4})$-$\log_{10}(\text{SFR})$ space, as determined by the overall best performing model. Points are coloured by $p$(SF), which is the probability of a source belonging to the SF-dominated component, as given by equation~(\ref{eq:PSF}). The various lines correspond to the best-fitting $\mu_{\text{SF}}$, $\mu_{\text{SF}} =  \log_{10}(\text{SFR}) = 0.790L' + 1.244 (1+z)^{0.122} -0.033M'$, with differing redshift and stellar mass values. The left panel shows the impact of varying redshift between $z=0.25$ and $z=3$ for a fixed $M_*=10^{10}M_\odot$, which results in a $\sim0.2$ dex difference in $\log_{10}(\text{SFR})$ for a given radio luminosity. The right panel shows the impact of varying $M_*$ over four orders of magnitude at a fixed $z=1$. For a fixed radio luminosity, the difference in $\log_{10}(\text{SFR})$ between $M_*=10^{8}-10^{12}M_\odot$ is $\sim0.13$ dex.}
    \label{fig:twopanel}
\end{figure*}

\subsection{The SFR-Radio Relation}
\begin{figure*}
    \centering
    \includegraphics[width=1\linewidth]{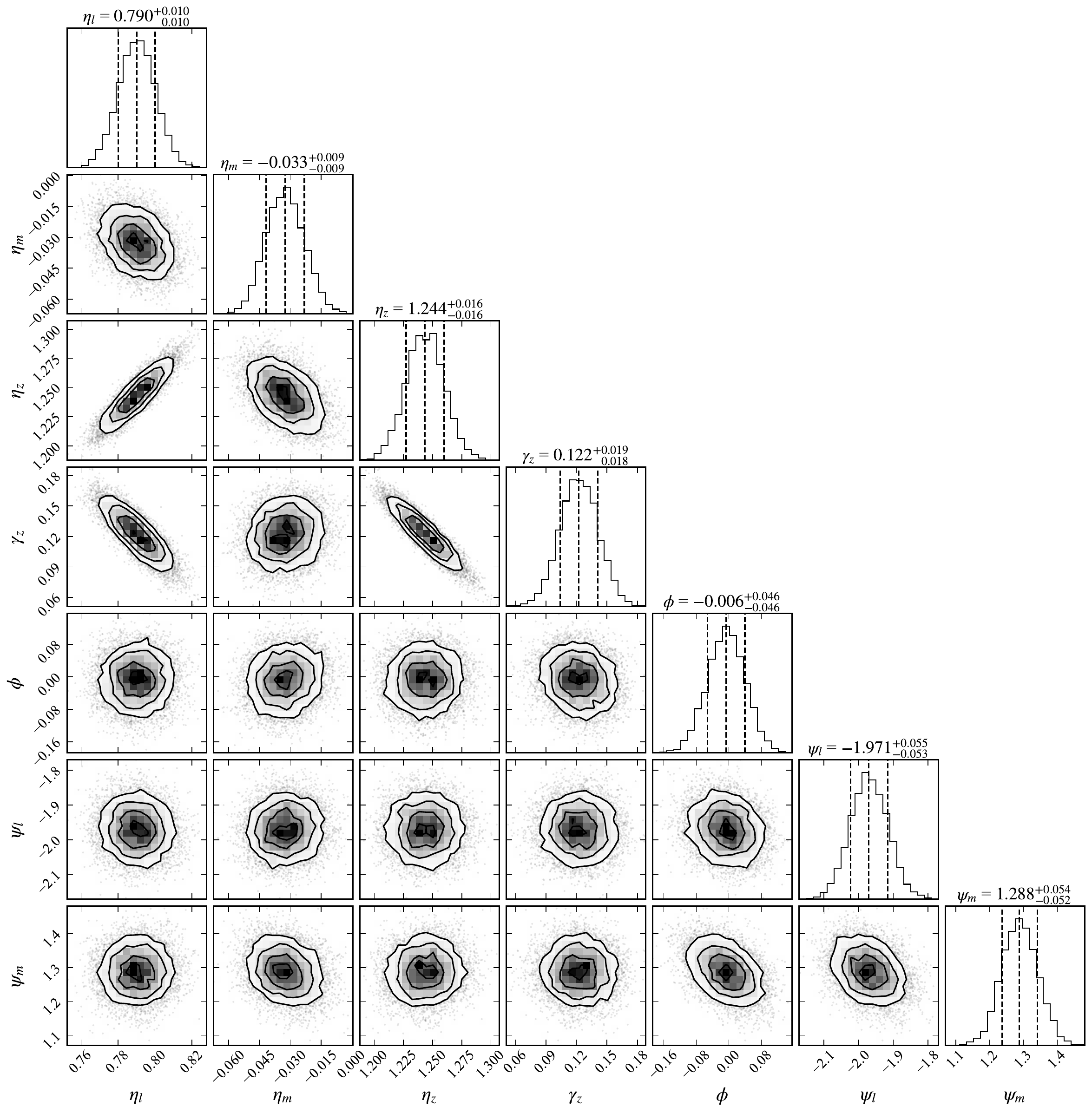}
    \caption{The posterior probability distributions of the free parameters in the optimal SFR-radio model,  $\mu_{\text{SF}}$[$M'$+$z$]; $f$ [$L'$+$M'$], as described by equations~(\ref{eq:bestfitmu}) and (\ref{eq:bestfitf}), ascertained by \texttt{NUTS} sampling. The median, 16 per cent and 84 per cent confidence intervals are stated.}
    \label{fig:bestfitcorner}
\end{figure*}

Our optimal model configuration is then sampled using \texttt{PyMC.sample}, yielding the identification of two components as shown in Fig.~\ref{fig:twopanel}. As expected, the model recovers a tight SF-dominated component, alongside a scattered background. The sampling returns posteriors for the parameters that can be used to derive a mean relation for the SFR-radio correlation, producing the following SFR-radio correlation:
\begin{equation}
\label{eq:bestfitmu}
    \mu_{\text{SF}} =  \log_{10}(\text{SFR}) = 0.790L' + 1.244 (1+z)^{0.122} -0.033M'.
\end{equation}
and shown in Fig.~\ref{fig:twopanel} as the solid black line.
The optimised parameters for equation~(\ref{eq:mu}) are presented in Table~\ref{tab:bestfit}, and the posterior distributions for those parameters are shown in Fig.~\ref{fig:bestfitcorner}. Most parameters are fit with little degeneracy, except between $\eta_{L},\eta_{z}$ and $\gamma_z$. Notably, the sampling does not find any significant degeneracy between the parameters describing stellar mass and redshift dependence. 

\begin{table*}
\centering
\caption{Posterior constraints on the best fitting model, $\mu_{\text{SF}}$[$M'$+$z$]; $f$ [$L'$+$M'$]. Values are medians, and the uncertainties stated are 68\% credible intervals.}
\label{tab:bestfit2}
\begin{tabular}{c|c|c|c|c|c|c|c|c|c}
\toprule
 $\eta_{\mathrm{l}}$ & $\eta_{\mathrm{m}}$ & $\eta_{z}$& $\gamma_{z}$  &$\phi$ &  $\psi_{\mathrm{l}}$& $\psi_{\mathrm{m}}$ & $\sigma_{\text{int}}$ & $\mu_{\mathrm{bkg}}$& $\sigma_{\mathrm{bkg}}$\\
\midrule
$0.790^{+0.010}_{-0.010}$ & $-0.033^{+0.009}_{-0.010}$ &$ 1.244 ^{+0.016}_{-0.016}$&$0.122^{+0.019}_{-0.019}$  & $-0.006^{+0.046}_{-0.046}$&$-1.971^{+0.055}_{-0.053}$  & $1.288^{+0.054}_{-0.052}$  & $0.178^{+0.004}_{-0.004}$  & $0.769^{+0.013}_{-0.013}$  & $0.762^{+0.009}_{-0.008}$\\
\bottomrule
\label{tab:bestfit}
\end{tabular}
\end{table*}

\begin{table}
\centering
\caption{Posterior constraints for the second and third best performing model configurations. Values are medians with 68\% credible intervals. Some variables are not present in both models, so are only quoted when relevant. It is important to note that model $\mu_{\text{SF}}$[$z$]; $f$[$L'$+$M'$] shows bimodal, highly degenerate posterior distributions, so the quoted medians and uncertainties should be treated with caution.  }
\label{tab:2n3}
\begin{tabular}{crrrr}
\toprule
Parameter & $\mu_{\text{SF}}$[$M'$+$z$]; $f$[$L'$+$z$+$M'$]  & $\mu_{\text{SF}}$[$z$]; $f$[$L'$+$M'$] \\
\midrule
\vspace{2pt}
$\eta_{\mathrm{l}}$
& $0.789\,{ ^{+0.010}_{-0.010}}$
& $0.776\,{ ^{+0.011}_{-0.012}}$
 \\
\vspace{2pt}
$\eta_{\mathrm{m}}$
& $-0.032\,{ ^{+0.009}_{-0.009}}$
& --  \\
\vspace{2pt}
$\eta_{z}$
& $1.241\,{ ^{+0.017}_{-0.017}}$
& $0.974\,{ ^{+0.765}_{-1.311}}$ 
\\
\vspace{2pt}
$\gamma_{z}$
& $0.126\,{ ^{+0.020}_{-0.020}}$
& $0.113\,{^{+0.100}_{-1.092}}$ 
\\
\vspace{2pt}
$\zeta$
& --
& $-0.094\,{ ^{+1.574}_{-0.627}}$ 
 \\
\vspace{2pt}
$\phi$
& $0.045\,{ ^{+0.096}_{-0.096}}$
& $0.320\,{ ^{+0.052}_{-0.052}}$ 
\\
\vspace{2pt}
$\psi_{\mathrm{l}}$
& $-2.022\,{ ^{+0.082}_{-0.074}}$
& $-2.077\,{^{+0.055}_{-0.055}}$
 \\
\vspace{2pt}
$\psi_{\mathrm{m}}$
& $1.277\,{^{+0.057}_{-0.054}}$
& $1.307\,{ ^{+0.054}_{-0.053}}$ 
 \\
\vspace{2pt}
$\psi_{z}$
& $-0.051\,{ ^{+0.080}_{-0.083}}$
& -- 
 \\
\vspace{2pt}
$\sigma_{\mathrm{int}}$
& $0.178\,{ ^{+0.004}_{-0.003}}$
& $0.180\,{ ^{+0.003}_{-0.004}}$
\\
\vspace{2pt}
$\mu_{\mathrm{bkg}}$
& $0.770\,{ ^{+0.013}_{-0.014}}$
& $0.837\,{ ^{+0.014}_{-0.013}}$ 
\\
\vspace{2pt}
$\sigma_{\mathrm{bkg}}$
& $0.762\,{ ^{+0.011}_{-0.008}}$
& $0.716\,{ ^{+0.009}_{-0.008}}$
\\
\bottomrule
\end{tabular}
\end{table}

Parameter uncertainties are given in Table \ref{tab:bestfit}. A key conclusion from our result is that a redshift evolution in the SFR-radio relationship is strongly preferred by the data. All of the highest scoring model configurations require a redshift term in the $\mu_{\text{SF}}$ relation, and in our best fitting solution the exponent on the $(1+z)$ term is 6.4$\sigma$ above zero. In the left panel of Fig.~\ref{fig:twopanel} we demonstrate the effect of varying redshift, showing that at higher redshifts, the same SFR is associated with lower radio emission. In our suite of tested models we included a multiplicative form, that would have allowed redshift to vary the slope of the relation rather than the intercept, but this was strongly disfavoured. On the other hand, the mass dependence found in Table \ref{tab:bestfit} is only 3.7$\sigma$ above zero and as can be seen in Fig.~\ref{fig:twopanel}, even an increase of multiple orders of magnitude in stellar mass yields virtually indistinguishable results. Indeed, a change in $M_*$ of four orders of magnitude produces only a 0.1 dex difference in $\log_{10}(\text{SFR})$ in our best fitting model, much less than even the intrinsic scatter in the relationship. Therefore, whilst the formal version of our derived SFR-radio includes a mass evolution term, in practise it could be safely omitted from calculation at little cost. For completeness, we also give the optimal parameter values for our second and third best performing models in Table \ref{tab:2n3}. We are confident that the observed evolutionary trend with redshift is not induced through an increasing mass incompleteness at higher redshifts, as our evolving background fraction acts to absorb any changing completeness in radio luminosity or stellar mass, as discussed in Section \ref{sec: pop evol}. Furthermore, even when the fitting is performed without a redshift term (i.e. as per $\mu_{\text{SF}}$[$M'$]; $f$[$L'$+$M'$], our fourth best ranking model in Table \ref{tab:model_comp}), a very mild mass dependence of $\eta_{\mathrm{m}} = -0.038^{+0.009}_{-0.010}$ is found, and instead $\eta_{\mathrm{l}}$ steepens to a value of $0.843^{+0.006}_{-0.007}$. Therefore, using our entirely data-driven methodology, we can conclude that the SFR-radio correlation shows very limited evolution with stellar mass.
Interestingly, if we apply a strict $\chi^2$ cut to our SED fitting results, the mass evolution term becomes weaker still, perhaps suggesting that the derived mass dependence of previous studies may be influenced by poorly constrained physical properties of sources, particularly if AGN have not been properly accounted for.

\subsection{Population Fraction Evolution}
\label{sec: pop evol}
Our \texttt{NUTS} sampling also reveals an optimal population fraction evolution described by:
\begin{equation}
\label{eq:bestfitf}
 f = \frac{1}{1+\exp\left[-(0.006 -1.971\times L'+1.288 \times M') \right]},
\end{equation}

\noindent where all parameters and their uncertainties are given in Table \ref{tab:bestfit}. 
The physical implication of this is that the fraction of SF-dominated sources decreases with increasing radio luminosity, yet increases with increasing stellar mass. The trend with radio luminosity corresponds to the highest radio luminosity sources being associated with powerful AGN activity, much more luminous than what can be attributed to star formation alone. The increasing fraction of SF-dominated sources at higher $M_*$ can be understood as our evolving fraction capturing the mass bias imposed by a flux limited sample. That is, our evolution of $f$ should not be strictly interpreted as though highest mass galaxies are physically preferentially SF-dominated, but instead that only those SF-dominated sources with relatively high $M_*$ can produce enough radio emission to meet the detection threshold. 
That being said, there is also the physical effect of the main sequence of star formation, where more massive galaxies have higher SFRs compared to lower mass galaxies \citep[][]{Noeske2007,Whitaker2014A, Johnston2015}. Clearly, the trends in $f$ should be carefully interpreted as a convolution of genuine underlying population trends and selection effects that the evolving background formulation acts to capture and prevent from impacting the overall SFR-radio correlation. It should equally be noted that the absolute values derived for our population fraction evolution (equation~(\ref{eq:bestfitf})), are highly affected by the survey flux limit, and will certainly vary if the analysis is repeated on a survey of a differing depth, by design.

\subsection{Redshift Evolution: Exploring the Possible Impact of Spectral Index Evolution}

There has been much debate in the literature about the existence of a redshift dependence in the SFR-radio relation, though most work has focussed on potential evolution in the IRRC, which is assumed to propagate through from the underlying SFR-radio relation-- though see \citet{molnar_non-linear_2021} for evidence that non-linearities in the IRRC drive observed dependencies. \citet[][]{magnelli_far-infraredradio_2015}, \citet{delhaize_vla-cosmos_2017}, and \cite{ocran_cosmic_2020} all find a mild, negative evolution in the (F)IRRC, while \citet{Algera2020}, \citet{DeZotti2024}, and \citet{Tabatabaei2025} find no redshift evolution in the range $1.5<z<3.5$, for example. In this work, we account for mass and redshift dependant selection effects through the use of our evolving SFG-AGN fraction, and recover a mild but statistically non-negligible evolution in the SFR-radio relation with redshift, though note that the number density of our sample is significantly reduced at $z>4$ compared to $0.5<z<2$, and therefore the conclusions we are able to draw are most robust at low to intermediate redshifts (e.g $z\lesssim4$).

There are multiple potential explanations for this evolution. Amongst those important to consider is the impact of inverse Compton (IC) scattering and synchrotron losses on the radio spectral index. 
Our sample spans a large redshift range and whilst IC losses from interactions with the cosmic microwave background (CMB) are minimal locally, they scale with the energy density of the CMB, which goes as $(1+z)^4$. \cite{murphy2009} demonstrates, for example, that this should become significant by $z\approx3$, and \cite{whittam2025} finds compelling evidence for a redshift evolution in 1.4~GHz flux density and luminosity at $3<z<5$ that is consistent with IC scattering, in line with previous findings by \citet{DeZotti2024}.  
In our primary analysis, we choose to take the standard literature approach of assuming the radio emission we observe in our sample is dominated at 1.4\,GHz by synchrotron emission, and can be treated with a constant $\alpha=-0.7$. However, any increase in electron cooling effects with redshift could quite reasonably lead to a radio spectral steepening, meaning our flux to luminosity conversion is underpredicting radio luminosities in sources at higher redshift. Therefore, a moderate positive redshift evolution would be needed to account for this in the data: much like we recover in our model. Motivated by this, we perform a simple test to ascertain the strength of spectral index evolution required to explain the derived redshift term in our SFR-radio relationship.

We move the calculation of radio luminosity to within the sampling procedure, parametrising the spectral index as $\alpha(z) = \alpha_0 + \eta_\alpha z$, where $\alpha_0 = -0.7$. The non-evolving case can therefore be recovered if the best fitting parameter of $\eta_\alpha$ is zero. The flux to luminosity conversion is therefore the same as equation~(\ref{eq:ref}), but with $\alpha \equiv\alpha(z)$. In our fitting of the SFR-radio relation, we use $\log_{10}(L_{1.4})$. Taking the base-10 logarithm of equation~(\ref{eq: lrad}), it can shown that the evolving spectral index case described above is mathematically equivalent to a linear additive redshift term, i.e.
\begin{equation}
\begin{gathered}
    \log_{10}(L_{1.4})[\alpha(z)] = \log_{10}(4\pi d_L^2 S_{1.4}  ) - (1+ \alpha_0 + \eta_\alpha z)\log_{10}(1+z) \\
=\log_{10}(L_{1.4})[\alpha_0]-\eta_\alpha z \log_{10}(1+z).
\end{gathered}
\end{equation} 
The sole difference between the implementation of an evolving spectral term within the radio luminosity calculation and a simple redshift term of the form $\eta_\alpha z \log_{10}(1+z)$ in $\mu_{\text{SF}}$ is whether the new calculated luminosity is included in the background fraction or not. We rerun the fitting procedure using both formulations, both of which require a constant intercept for normalisation purposes. The value of $\eta_\alpha$ in both cases is found to be $-0.068^{+0.025}_{-0.024}$. The other model parameters change slightly in comparison to the best fitting model from the primary analysis, and the optimised $\mu_{\text{SF}}$ is found to be \begin{equation}
\begin{gathered}
   \log_{10}(\text{SFR}) = 0.830 L’[\alpha(z)] - 0.037M’ + 1.329\\\
     \alpha[z] = -0.7 -0.068z
\end{gathered}
\end{equation}
or equivalently, 
\begin{equation}
 \log_{10}(\text{SFR}) = 0.830 L’[\alpha_0] - 0.037M’ + 0.056z\log_{10}(1+z) + 1.329.
\end{equation}

\begin{table*}
\centering
\caption{Posterior constraints on the model $\mu_{\text{SF}}$[$M'$+$z$]; $f$ [$L'$+$M'$] with an evolving spectral index, $\alpha = \alpha_0 - \eta_\alpha z$. Values are medians, and the uncertainties stated are 68\% credible intervals.}
\label{tab:specfit}
\begin{tabular}{c|c|c|c|c|c|c|c|c|c}
\toprule
$\eta_{\alpha}$& 
$\eta_{\mathrm{l}}$ & 
$\eta_{\mathrm{m}}$ & 
$\zeta$ &$\phi$ &  
$\psi_{\mathrm{l}}$&
$\psi_{\mathrm{m}}$ &
$\sigma_{\text{int}}$ &
$\mu_{\mathrm{bkg}}$& 
$\sigma_{\mathrm{bkg}}$\\
\midrule
$-0.068^{+0.025}_{-0.024}$ & 
$0.830^{+0.008}_{-0.008}$
& $-0.037^{+0.009}_{-0.009}$& 
$1.329^{+0.009}_{-0.009}$&
$0.059^{+0.047}_{-0.047}$ &
$-1.929^{+0.058}_{-0.057}$  &
 $1.274^{+0.054}_{-0.051}$ & 

$0.180^{+0.004}_{-0.004}$  & 
$0.764^{+0.009}_{-0.009}$&
$0.768^{+0.013}_{-0.013}$\\
\bottomrule
\end{tabular}
\end{table*}
\begin{figure}
    \centering
    \includegraphics[width=1\linewidth]{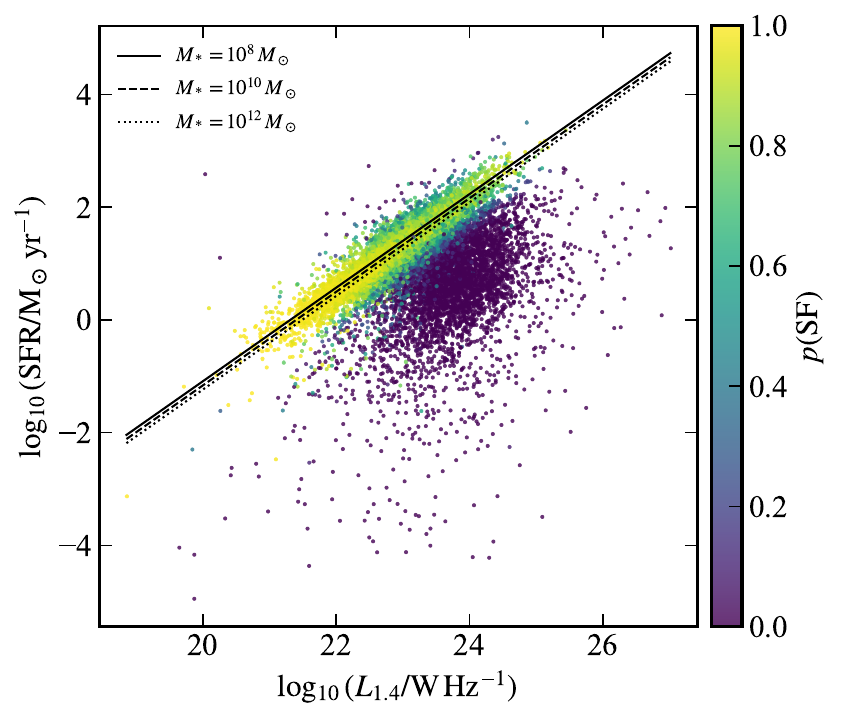}
    \caption{The distribution of sources in $\log_{10}(L_{1.4})$-$\log_{10}(\text{SFR})$ space, with an evolving $\alpha $ factor included in the calculation of $\log_{10}(L_{1.4})$, of the form $\alpha = -0.7 -0.068z$. Points are coloured by $p$(SF), which is the probability of a source belonging to the SF-dominated component, as given by equation~(\ref{eq:PSF}). The lines correspond to the SFR-radio relation, $ \log_{10}(\text{SFR}) = 0.830 L’[\alpha(z)] - 0.037M’ + 1.329$, evaluated at different stellar masses. The 4 order of magnitude change in stellar mass corresponds to a change of 0.12 dex in SFR. }
    \label{fig:spec}
\end{figure}
\begin{figure*}
    \centering
    \includegraphics[width=1\linewidth]{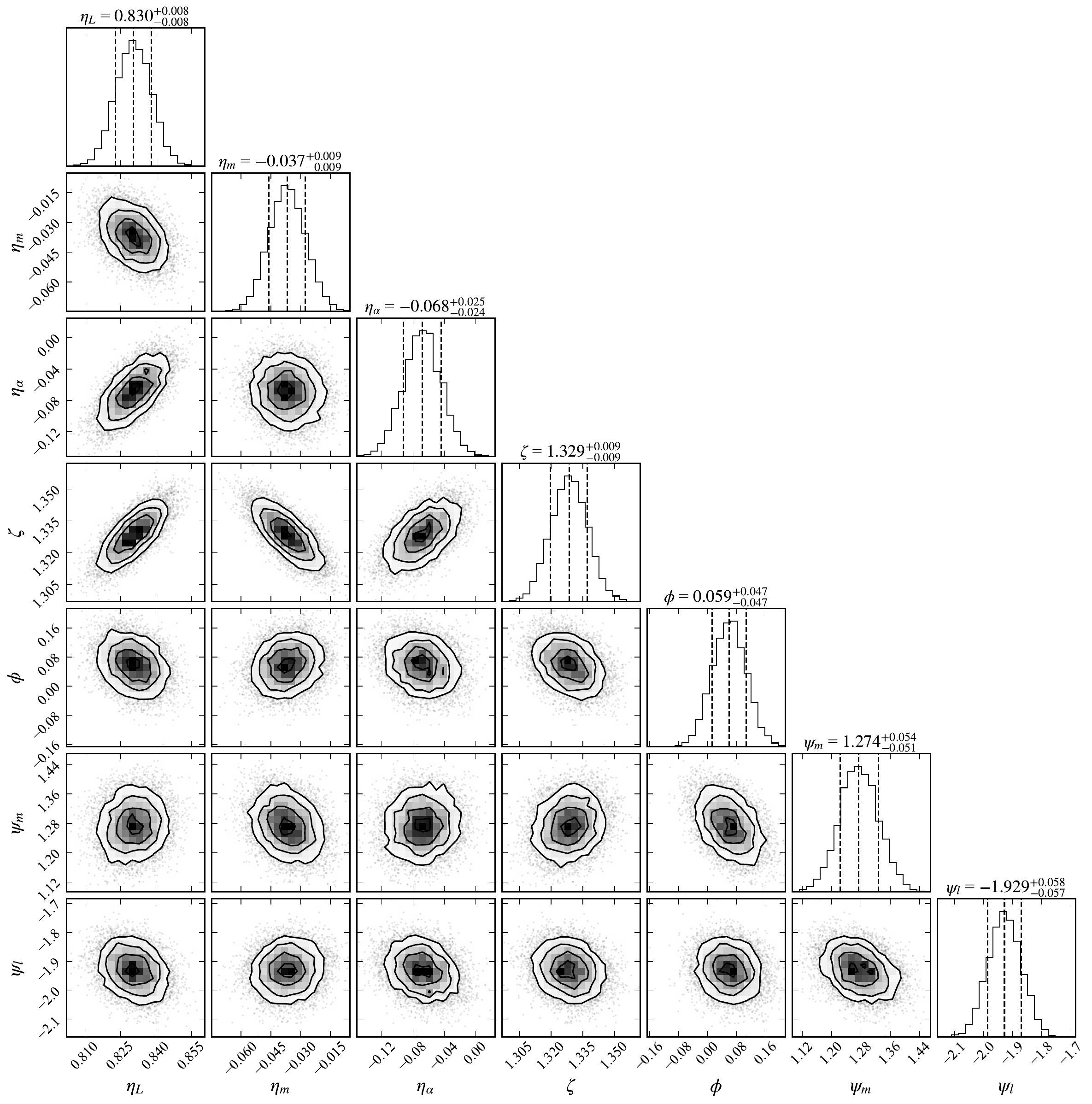}
    \caption{The corner plot describing the best fitting combination of parameters for the evolving spectral index model, as ascertained by the \texttt{NUTS} sampling. The median, 16 per cent and 84 per cent confidence intervals are stated.}
    \label{fig:speccorner}
\end{figure*}

These results are presented in Fig.~\ref{fig:spec}, with the values and uncertainties presented in Table \ref{tab:specfit} and the posterior distributions shown in Fig.~\ref{fig:speccorner}. The WAIC and PSIS-LOO-CV would place these evolving spectral index models as the 4th and 5th best  overall, with a best case difference of $\Delta$ELPD$_{\text{WAIC}} = 15.88 \pm 5.11 $ and $\Delta$ELPD$_{\text{LOO}} = 15.90 \pm 5.11 $, compared to the highest ranking model. We do not set $a_0$ as a free parameter in our test, as our current single frequency data is insufficient to constrain it, and note that setting it to a different (reasonable) value causes a small change to the inferred evolution, as might be expected. For example, the case of $\alpha_0=-0.75$ gives the evolution $\alpha[z] = -0.7 - 0.057^{+0.022}_{-0.023}z$, which is weaker but consistent within errors to the $\alpha_0 = -0.7$ case.\\

Recent work by \citet{2026arXiv260527149H} defined a unified `fundamental plane' of the SFR-radio relation for both individual galaxies and kiloparsec-scale spatially resolved regions by introducing an additional linear spectral index term to the form of equation~(\ref{eq:simple}). They attribute the dependence to a steepening of measured spectral index in sources that undergo more significant synchrotron losses in their CRs, alongside reduced radio luminosity in sources with more efficiently escaping CRs. 
In our case, we choose not to speculate as to whether or not a spectral index evolution is the genuine cause of the observed redshift dependence, but note that our value of $\eta_\alpha=-0.068$ corresponds to a spectral index of $\alpha(z=0)=-0.7$, $\alpha(z=1)=-0.768$ and $\alpha(z=2)=-0.836$, values which would not be especially extreme. An alternative explanation beyond a physical spectral index evolution would be if the population of radio detected galaxies had a curved spectrum, rather than one that could be described by a single spectral index, or even just an average spectral index greater than the assumed $\alpha=-0.7$. At higher redshift, the observed 1.4$\,$GHz radiation has its origins at increasingly high frequencies, scaling as $\propto1.4(1+z)$. Therefore, using a single $\alpha=-0.7$ value would increasingly underestimate the rest frame luminosity as redshift increased -- inducing an observed redshift evolution in the sample. Furthermore, even if a single $\alpha$ value were to adequately describe the average radio spectrum of our sources, as a higher rest-frame frequency is probed, the fundamental relationship between SFR and $L_{\nu, \text{emit}}$ may not be constant. An example of this in the local universe is shown in \citet{2017ApJ...836..185T}. Equally, it is possible that there is a breakdown of the assumptions underpinning the SFR-radio relation at higher redshifts -- for example, the reduction in radio luminosity with increasing redshift for a fixed SFR and $M_*$ could point towards a potential disconnect in the calorimetry models that link the two properties. This could occur due to galaxies generally being more compact at earlier epochs, resulting in a greater fraction of CR electrons being able to escape from the source without interacting with magnetic field lines, thus producing weaker observable synchrotron radiation.

Fundamentally, it is difficult to disentangle which of these possible causes of an observed redshift evolution is ultimately responsible for the relationship we find; however, there is clearly good reason to expect that the SFR-radio correlation should vary with redshift in one manner or another. Whether the apparent redshift evolution of our SFR-radio relationship is a result of an evolving spectral index; an innate link between spectral index, CR escape efficiency and radio luminosity; curved radio spectra, or perhaps some alternative cause will rapidly enter the realm of testability with the MIGHTEE S-band data (Hale et al. in prep.) and (Thykkathu et al. in prep.) over the MIGHTEE-DR1 fields. 


\subsection{Testing for Dependence on Selection Effects}
\label{sec:select}

The main selection effect that may impact our results if improperly treated is the Malmquist bias -- for example, is our measured correlation between $L_{1.4}$ and SFR simply a result of comparing luminosities from two flux limited samples? The fact that we fit for both redshift, mass and luminosity dependencies in the correlation between SFR and radio luminosity, somewhat mitigates the impact of Malmquist bias. Furthermore, our parent sample of optical and near-infrared galaxies is stellar mass limited, which although correlated with SFR through the star-formation main sequence \citep[][]{Noeske2007,Whitaker2014A, Johnston2015}, is not directly impacted by the optical/near-infrared flux limit.
We also have a relatively large baseline in both radio luminosity and redshift, allowing any SFR--radio relation to be constrained by the data at low redshift, whereas the redshift evolution constraints are obviously more influenced by the high-luminosity SFGs that we detect across a broad redshift range. Similarly, any dependence of the SFR--$L_{1.4}$ relation on stellar mass is largely constrained by the same lower-redshift populations where we have the largest baseline in stellar mass from $\sim10^8 - 10^{11}$\,M$_{\odot}$ (Fig.~\ref{fig:radz}).
\begin{figure}
    \centering
    \includegraphics[width=1\linewidth]{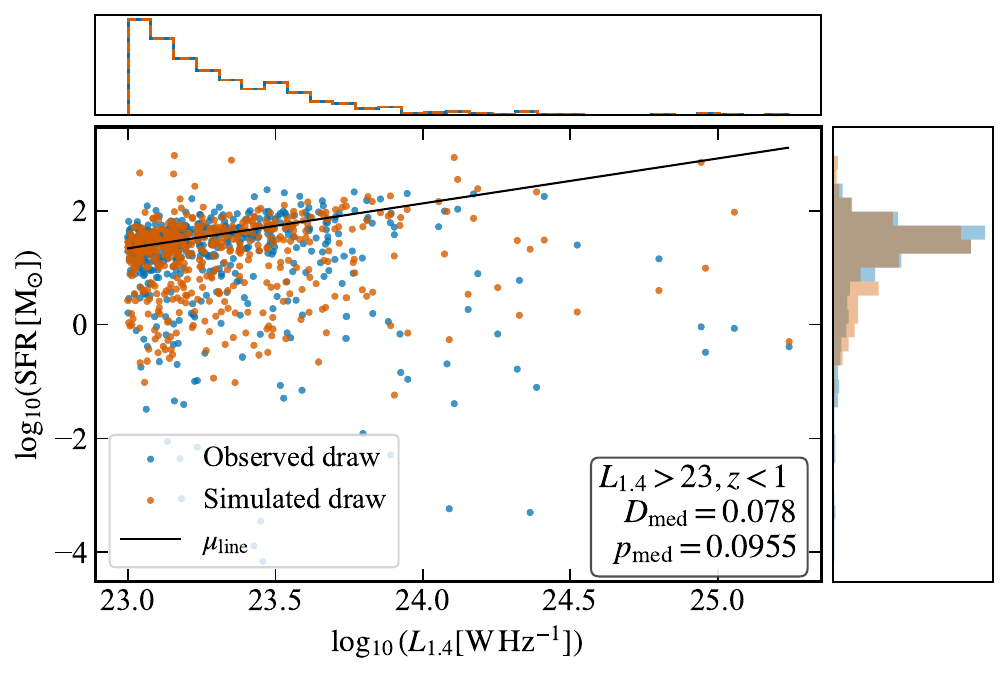}
    \includegraphics[width=1\linewidth]{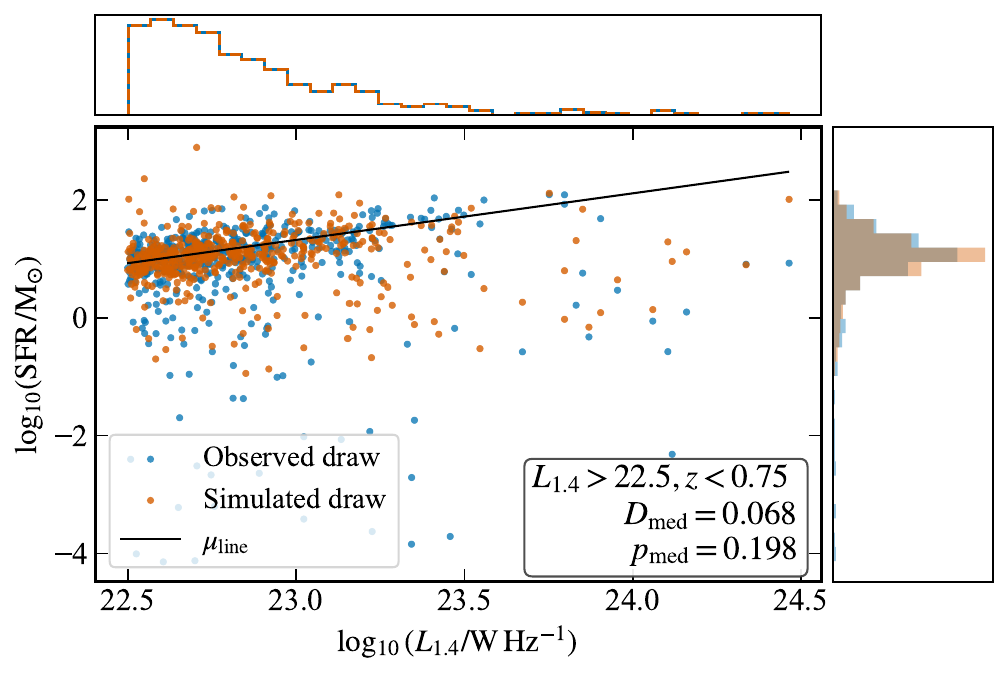}
 
    \caption{In SFR-$L_{1.4}$ space, a realisation of distribution of observed sources (blue)-- where their SFR value has been randomly selected from their posterior-- and simulated sources (orange), calculated from observed $L_{1.4}$, $z$ and $M_*$ values combined with the posterior probability distributions of the fitted parameters in our derived SFR-radio correlation. The line on each plot shows the median best fitting values of the SFR-radio correlation, i.e. equation~(\ref{eq:bestfitmu}). A KS test has been performed on the SFR distributions and the median KS statistic ($D_{\text{med}}$) and p-val ($p_{\text{med}})$ are quoted. Both panels represent a different volume limited sample: the top has been restricted to $L_{1.4} >23, z<1$ and the bottom $L_{1.4} >22.5, z<0.75$.}
    \label{fig:malmquest}
    
\end{figure}

To test the impact of Malmquist bias we limit our sample to a region in the $L_{1.4}-z$ plane where we can expect that observations are volume limited, i.e. complete to the lower $L_{1.4}$ limit, such that if the same inferred SFR-radio correlation holds then it can be safely separated from Malmquist bias.  We restrict our catalogue to sources in a defined region of $L_{1.4}-z$ space, indicated in Fig. \ref{fig:radz}, and draw a bootstrap sample of 500 sources from this volume-limited catalogue. The choice is made to bootstrap to a relatively small number in order to minimise the impact of sample size on our eventual population statistics, but with many realisations to ensure robustness. For each source, we randomly sample from their SFR posterior distributions from the \texttt{GRAHSP} SED fitting to produce an observed benchmark. A single full posterior sample of the fitted model parameters is randomly selected from the chain, and is then combined with the observed $L_{1.4}$, $z$ and $M_*$ values of the sources to calculate a probability of belonging to either component, given by $f$, as in equation~(\ref{eq:bestfitf}). A Bernoulli draw based on this probability then dictates whether an individual source will have its SFR calculated using
equation~(\ref{eq:bestfitmu}) (i.e. it belongs to the SFR-radio correlation) or if it is assigned to the Gaussian background with mean $\mu_{\text{bkg}}$ and $\sigma_{\text{bkg}}$ (values given in Table \ref{tab:bestfit}). The simulated and observed catalogues are compared with a Kolmogorov--Smirnov (KS) test from \textsc{Scipy} \citep{scipy}. We choose a KS-test over the commonly used Anderson-Darling (AD) test, as AD tests are more sensitive to differences in the tail of distributions compared to a KS-test, and we are primarily concerned with how well the SF-dominated relationship is recovered, rather than the background. This procedure is then repeated 5000 times, across many different posterior realisations, and thus yields a distribution of KS statistics and p-values through which we can ascertain how well the equations~(\ref{eq:bestfitmu}) and (\ref{eq:bestfitf}) reproduce the observed population within a volume limited sample. 

We perform the test above on both the case of $\log_{10}(L_{1.4} ) > 23, z<1$ and $\log_{10}(L_{1.4}) >22.5, z<0.75$, which are both indicated in Fig.~\ref{fig:radz}. The results are shown in Fig.~\ref{fig:malmquest}, and the median KS test statistics are 0.078 for $\log_{10}(L_{1.4} ) > 23, z<1$ and 0.068 for $\log_{10}(L_{1.4}) >22.5, z<0.75$, and the median p-values are 0.096 and 0.198 respectively. Therefore, in both cases, the simulated samples can be taken as statistically indistinguishable from the observed samples. Furthermore, as can be seen from the alignment of $\mu_{\text{SF}}$ as described in equation~(\ref{eq:bestfitmu}) and in Fig.~\ref{fig:malmquest}, our derived relation holds equally well in a complete, volume limited sample, and we can conclude that it is unlikely to be a result of flux-related selection bias.

Another potential issue that could affect our results is an overweighting of high redshift sources, both as a consequence of ($1+z$) not being proportional to comoving volume, and as the uncertainty on the intrinsic assumed spectral index increases. We verify that our results are insensitive to an imposed lower redshift limit ($z<2$) by rerunning the fit for that subsample, finding parameter values $\eta_L=0.794\pm0.010$, $\eta_m=-0.035\pm0.000$, $\eta_z=1.258\pm0.017$, $\gamma_z=0.103\pm0.021$, $\phi=0.221^{+0.050}_{-0.051}$, $\psi_m=1.244\pm{0.061}$, $\psi_L=-2.083^{+0.055}_{-0.059}$ and $\sigma_{\text{int}}=0.181^{+0.04}_{-0.03}$. All free parameters are reproduced within the confidence interval of the value derived from the full sample, importantly including those determining the redshift evolution, and the conclusions of this paper are unchanged regardless of which redshift range is included in the analysis.

\section{Conclusions}

We have presented a new, improved method for determining the star formation rate -- $L_{1.4}$ correlation in star-forming galaxies. The principal strengths of this method include the flexible treatment of contaminating AGN (rather than needing to rely on binary cuts), a full inclusion of SFR posterior probabilities in the calculation, and a rigorous testing of the optimal functional form for the relationship. By applying this method to a combination of deep MIGHTEE radio flux densities and SFR estimates derived from state-of-the-art SED fitting with the new code \texttt{GRAHSP}, we obtain the relationship
\begin{equation*}
    \mu_{\text{SF}} =  \log_{10}(\text{SFR}) = 0.79L' + 1.244 (1+z)^{0.122} -0.033M',
\end{equation*}
where $L' = \log _{10}(L_{1.4}/\text{W\,Hz}^{-1})-23$ and $M' =\log _{10}(M_*/M_{\odot})-10 $. 
A key result is that we find a mild but statistically significant positive redshift evolution, such that at a higher redshift there is less radio luminosity associated with the same SFR. We find a minimal stellar mass evolution, much weaker than previous works based on the IRRC. 

The redshift evolution that we recover can reasonably be explained by a mild spectral index evolution in redshift of the form  $\alpha = -0.7 -0.068z$, and an associated SFR-radio correlation of $\log_{10}(\text{SFR}) = 0.830 L’[\alpha(z)] - 0.037M’ + 1.329$. We do not know if this scenario represents physical reality, but it will be directly testable with matched-resolution imaging of the MIGHTEE fields in both L- and more recent S-band that will become available in the near-future (Hale et al. in prep.; Thykkathu et al. in prep.). We also show that our derived correlation is unlikely to be a direct result of Malmquist bias.

Previous studies have found a much stronger dependence of the SFR -- radio correlation on the stellar mass \citep[e.g.][]{2018MNRAS.475.3010G, delvecchio_infrared-radio_2021, Smith2021} than found in this work. Although it is difficult to assess the root cause of this difference, it is likely due to divergent approaches to the problem, specifically how different studies deal with AGN contamination is likely to be a critical factor. The mass dependencies found previously clearly show that the radio luminosity increases with the mass of the galaxy, and the main constraints on this are from galaxies with $M_* > 10^{10}$\,M$_{\odot}$. As highlighted by \cite{Smith2021}, it is difficult to rule out AGN contamination in this regime with methods that apply cuts based on distance from an (assumed) underlying correlation prior to measuring the correlation itself. Furthermore, works that obtain IR luminosities from SED fitting have rarely included appropriate AGN components in their modelling, further adding to the issue of AGN contamination in their results. On the other hand, our method is purely driven by the data itself, using a statistical approach that marginalises over the full per-source SFR posterior distribution  derived from \texttt{GRAHSP} SED fitting, where AGN contributions are flexibly accounted for, coupled with a hierarchical Bayesian model that accounts for an evolving background population (AGN). We therefore suggest that the strong mass dependence is at least partially an artefact of the initial sample selection of SFGs (or removal of AGN from the sample). Furthermore, we directly derive the SFR-radio relationship, without using an intermediate scaling relation such as the IRRC. If there is a stellar mass dependence in the way that the IR and radio emission links in relation to SF activity, then using the IRRC could artificially induce a stellar mass evolution in the resulting SFR-radio relationship. 

The relation from this work can be widely applied to studies investigating the evolution of the star-formation history of the Universe using radio data. However, we caution that applying it naively to the general radio population where traditional cuts have been made to remove AGN will continue to result in uncertain levels of AGN contamination in any derived quantities. As such, we may need to move towards a new framework, such as implemented in this work, to obtain more accurate measurements of the star formation in galaxies traced via radio emission. Furthermore, given that AGN emission can contaminate across the whole of the electromagnetic spectrum, such an approach may also be required for other tracers of star formation. 
\section*{Acknowledgements} 
CLJ and JHM acknowledge funding from a Royal Society University Research Fellowship (URF\textbackslash R1\textbackslash221062). CLJ gratefully acknowledges support from the Balliol College Dervorguilla Scholarship. MJJ, IHW and CLH acknowledge support from the Hintze Family Charitable Foundation through the Oxford Hintze Centre for Astrophysical Surveys. MJJ, NS, RGV and SLJ acknowledge funding from a UKRI Frontiers Research Grant [EP/X026639/1], and MJJ acknowledges the support of the Science and Technology Facilities Council (STFC) consolidated grant [ST/W000903/1]. JH acknowledges funding from STFC [grant code ST/Y509474/1]. CLH acknowledges support from STFC through grant ST/Y000951/1. DJBS acknowledges support from STFC under grant ST/Y001028/1, and from the Leverhulme Trust via Research Project Grant RPG-2025-078. We thank Nick Seymour, Bohan Yue, and Fatemeh Tabatabaei for their helpful comments and suggestions, and the anonymous reviewer for their useful feedback.

The MeerKAT telescope is operated by the South African Radio Astronomy Observatory, which is a facility of the National Research Foundation, an agency of the Department of Science and Innovation. We acknowledge the use of the ilifu cloud computing facility – www.ilifu.ac.za, a partnership between the University of Cape Town, the University of the Western Cape, Stellenbosch University, Sol Plaatje University and the Cape Peninsula University of Technology. The Ilifu facility is supported by contributions from the Inter-University Institute for Data Intensive Astronomy (IDIA – a partnership between the University of Cape Town, the University of Pretoria and the University of the Western Cape, the Computational Biology division at UCT and the Data Intensive Research Initiative of South Africa (DIRISA). The authors acknowledge the Centre for High Performance Computing (CHPC), South Africa, for providing computational resources to this research project.
This work is based on data products from observations made with ESO Telescopes at the La Silla Paranal Observatory under ESO programme ID 179.A-2005 (Ultra-VISTA) and ID 179.A-2006 (VIDEO) and on data products produced by CALET and the Cambridge Astronomy Survey Unit on behalf of the Ultra-VISTA and VIDEO consortia.
The Hyper Suprime-Cam (HSC) collaboration includes the astronomical communities of Japan and Taiwan, and Princeton University. The HSC instrumentation and software were developed by the National Astronomical Observatory of Japan (NAOJ), the Kavli Institute for the Physics and Mathematics of the Universe (Kavli IPMU), the University of Tokyo, the High Energy Accelerator Research Organization (KEK), the Academia Sinica Institute for Astronomy and Astrophysics in Taiwan (ASIAA), and Princeton University. Funding was contributed by the FIRST program from Japanese Cabinet Office, the Ministry of Education, Culture, Sports, Science and Technology (MEXT), the Japan Society for the Promotion of Science (JSPS), Japan Science and Technology Agency (JST), the Toray Science Foundation, NAOJ, Kavli IPMU, KEK, ASIAA, and Princeton University.
Based on observations obtained with MegaPrime/MegaCam, a joint project of CFHT and CEA/IRFU, at the Canada-France-Hawaii Telescope (CFHT) which is operated by the National Research Council (NRC) of Canada, the Institut National des Science de l’Univers of the Centre National de la Recherche Scientifique (CNRS) of France, and the University of Hawaii. This work is based in part on data products produced at Terapix available at the Canadian Astronomy Data Centre as part of the Canada-France-Hawaii Telescope Legacy Survey, a collaborative project of NRC and CNRS. 
\section*{Data Availability}
All observational data used in this paper are publicly available. See \citet{haledr1} for the MIGHTEE radio data, and \citet{mighteehosts} for the cross-matched catalogue. The \texttt{GRAHSP} SED fits and derived SFRs used in this paper will be released in a subsequent paper (Jackson et al. in prep.). Additional derived data products are available on reasonable request.



\bibliographystyle{mnras}
\bibliography{mybib} 





\bsp	
\label{lastpage}
\end{document}